%% file: main.tex
\pdfoutput=1

\documentclass{article}

\usepackage{microtype}
\usepackage{graphicx}
\usepackage{subfigure}
\usepackage{booktabs} 

\usepackage{amsmath}
\usepackage{amssymb}
\usepackage{xcolor}
\usepackage[normalem]{ulem}   

\usepackage{hyperref}

\usepackage[accepted]{mlsys2025}
\makeatletter\renewcommand{\Notice@String}{Preprint.}\makeatother

\usepackage{comment}

\newcommand{\Description}[1]{}

\newcommand{\fixme}[1]{\textcolor{red}{#1}}

\newif\ifshowchanges
\showchangestrue

\ifshowchanges
  
  \newcommand{\del}[1]{\textcolor{red}{\sout{#1}}}
\else
  
  \newcommand{\del}[1]{}
\fi

\mlsystitlerunning{HE-LRM: Encrypted Deep Learning Recommendation Models using Fully Homomorphic Encryption}

\begin{document}

\twocolumn[
\mlsystitle{HE-LRM: Encrypted Deep Learning Recommendation Models using Fully Homomorphic Encryption}

\mlsyssetsymbol{equal}{*}

\begin{mlsysauthorlist}
\mlsysauthor{Akshath Mahajan}{equal,nyu}
\mlsysauthor{Karthik Garimella}{equal,nyu}
\mlsysauthor{Austin Ebel}{nyu}
\mlsysauthor{Gabrielle De Micheli}{lg}
\mlsysauthor{Brandon Reagen}{nyu}
\end{mlsysauthorlist}

\mlsysaffiliation{nyu}{New York University}
\mlsysaffiliation{lg}{LG Corporation}

\mlsyscorrespondingauthor{Akshath Mahajan}{avm6288@nyu.edu}

\vskip 0.3in

\input{00_abstract.tex}
]

\printAffiliationsAndNotice{\mlsysEqualContribution}

\input{01_introduction.tex}

\section{Background}

\subsection{CKKS High-Level Overview}\label{subsec:fhe}
CKKS is an FHE scheme that enables approximate computation directly on encrypted vectors without requiring the secret key. It encrypts vectors of a fixed power-of-two length (typically $n=2^{15}$ elements, or \emph{slots}) and supports SIMD addition, SIMD multiplication, and cyclic rotation. For addition and multiplication, one operand may be unencrypted, but the output remains a ciphertext. CKKS does not natively support indexing by an encrypted value; embedding lookups must therefore be expressed using these primitives.

CKKS represents values at an enlarged scale to preserve precision.
Multiplication increases this scale, and a \emph{rescaling} operation reduces it while consuming part of the ciphertext's computational budget.
A ciphertext's \emph{multiplicative level} counts its remaining rescaling steps.
Each rescaling consumes one level; when this budget is exhausted, an expensive \emph{bootstrapping} operation restores levels to enable further computation.
Linear transformations can be evaluated in one level using the baby-step giant-step (BSGS) method, while activation functions such as ReLU are approximated by polynomials that consume several levels~\cite{orion}.
The depth of these approximations often necessitates bootstrapping during neural inference.
Appendix~\ref{subsec:appendix_ckks} provides the underlying CKKS details.

Encrypted operations are both compute and memory intensive. Ciphertext-ciphertext multiplication and rotation require costly \emph{key-switching} operations. Ciphertexts, encoded plaintext weights, and evaluation keys also occupy substantial memory, making both the number of operations and their working set important to performance. Appendix~\ref{subsec:appendix_ckks} details these system-level costs.

\subsection{Deep Learning Recommendation Models}
\label{subsec:dlrm}

Deep Learning Recommendation Models (DLRMs) \cite{naumov2019deep} are neural network architectures that accept as input both continuous (dense) and categorical (sparse) features and output probabilities for recommendation systems. 
The model outputs a probability corresponding to the likelihood of accepting a particular recommendation. 
A \textit{bottom MLP} of linear layers with ReLU activations processes the dense features, and each sparse feature is mapped to a dense vector by its own embedding table.
For industry-scale DLRMs, these tables can contain millions of rows and consume GiBs to TiBs of memory \cite{yin2021tt}.
An interaction operation then combines the bottom MLP output with the embedding vectors, either through pairwise dot products or by concatenation.
In this work, we opt for the latter.
A \textit{top MLP} maps the combined vector to a single logit, followed by a sigmoid.
In our use-case, all inputs to the DLRM (both dense and sparse) must remain private.

\subsection{Threat Model}
\label{subsec:threat}
We assume a semi-honest threat model ~\cite{pragmaticmpc} in which an adversary faithfully takes part in the private inference protocol but may try to learn additional information from the messages they receive (also assumed in most prior works in private inference \cite{gazelle, delphi, orion}).
Moreover, it is assumed that the client \textit{knows the dimensionality of the input} to the neural network for which they query. In this paper, we make this assumption explicit given that the inputs are no longer only dense features but also the sparse features used for the embedding tables. To this end, we assume that the client knows the input size to the Bottom MLP and the sizes (number of rows) of each embedding table. Crucially, the client does not know the embedding dimension (number of columns) or parameters of the embedding tables, the interaction method, or the parameters of the MLPs. 


Our compressed embedding strategy adheres to this threat model. In contrast, prior works with embeddings either leak the learned embedding tables to the client~\cite{nexus,thor,zhang2025moai} or reveal correlations between embedding vectors~\cite{kim2024privacypreserving}. Because embedding tables are \textit{weights} learned during training, they are intellectual property of the model owner and leaking the tables or their correlations leaks information about the model itself.

\section{The Private Embedding Lookup Problem}
\label{sec:private-embedding-lookup}

We detail embedding lookups in CKKS for downstream FHE computations like DLRM inference.
Embedding tables are parameterized as dense matrices $E\in\mathbb{R}^{k\times d}$ where $k$ is the number of possible categories for a given categorical feature and $d$ is the embedding dimension.
Categorical values map to indexes $i\in\mathbb{Z}_k$, which we identify with $\{0,\ldots,k-1\}$, and the embedding layer returns the corresponding row $E[i]\in\mathbb{R}^d$.
In the cleartext setting, this lookup is a cheap memory access.
Under CKKS, however, indexing is not natively supported; computation is expressed through SIMD arithmetic and slot permutations.
Therefore, embedding lookup must be reformulated as an arithmetic computation over encrypted data.

A straightforward solution is to treat embedding lookup as a linear transformation by using an encrypted one-hot encoded index.
We define the one-hot encoding function as $\operatorname{OHE}_k:\mathbb{Z}_k\to\{0,1\}^k$, with a column-vector output whose coordinate $a$ is
\[
  \bigl(\operatorname{OHE}_k(i)\bigr)_a
  =\begin{cases}
    1 & \text{if } a=i,\\
    0 & \text{otherwise},
  \end{cases}
  \qquad 0\leq a<k.
\]
Then, by encrypting this basis vector $s_i=\operatorname{OHE}_k(i)$ under FHE, the server can perform the matrix-vector multiplication $E^\top s_i$ homomorphically.
This computation returns the selected embedding as a column vector, up to CKKS approximation error.

This approach decomposes private embedding lookups into two subproblems: (1) constructing the encrypted one-hot selector and (2) applying that selector to the server-owned embedding table.
We examine each of the aforementioned subproblems, and also consider how an alternative lookup basis can simplify encrypted vector construction.

\subsection{Constructing the Encrypted Selector}
\label{sec:selector-construction}

\noindent\textbf{Client-Side Selector Construction.}
The most direct way to construct the encrypted selector is for the client to perform the one-hot encoding locally before encryption.
Given a categorical index $i\in\mathbb{Z}_k$, the client computes $s_i=\operatorname{OHE}_k(i)\in\{0,1\}^k$ and sends its CKKS encryption to the server.
This requires no homomorphic selector construction and consumes no multiplicative levels. The server can use the received selector directly for the embedding lookup.

The client only needs to know the number of rows $k$ in the embedding table to construct the selector, which holds within our threat model (see Section~\ref{subsec:threat}).
Thus, under the security of CKKS, the server learns neither the selected index~$i$ nor which embedding row was returned.

The main drawback for client-side one-hot encoding is the low slot utilization.
A raw one-hot selector for a table with $k$ rows requires $k$ slots, 
of which exactly one is non-zero. Letting $n$ denote the number of slots in a  ciphertext, for large tables with $k > n$ the selector spans many ciphertexts and can induce substantial communication overhead.

\noindent\textbf{Server-Side Selector Construction.}
An alternative 
is to shift the selector construction to the server.
The client encrypts the categorical index itself and sends $\operatorname{Enc}(i)$ to the server.
The server homomorphically expands this encrypted index into an encrypted one-hot selector, which
at a high level requires approximating an equality test between the private index $i$ and each table row $a\in\mathbb{Z}_k$.

This reduces input communication at the cost of server computation.
Equality evaluation consumes multiplicative depth and may require bootstrapping, depending on the table size, precision, and FHE parameters.
The encrypted index need not be expanded into a one-hot selector, Section~\ref{sec:ive} describes an alternative basis: Independent Vector Evaluation. 


\subsection{Selecting the Embedding Row}
\label{sec:selecting-embedding-row}

We now detail two approaches to extracting the embedding row given the encrypted one-hot selector.
Both evaluate the same linear map but organize ciphertexts and rotations differently, exposing different tradeoffs under CKKS.
Assume $k,d\leq n$, so the selector and output each fit in one ciphertext.
Larger tables can be partitioned into blocks.

\noindent\textbf{Matrix-Vector Lookup.}
Once an encrypted selector vector is ready, 
the most direct formulation is a dense matrix-vector product.
For the one-hot selector $s_i$, the desired embedding row is given by $y=E^\top s_i$,
where $y\in\mathbb{R}^d$ is the selected embedding vector.

\begin{figure}[t]
  \centering
  \includegraphics[width=\linewidth,trim=0 18 0 0,clip]{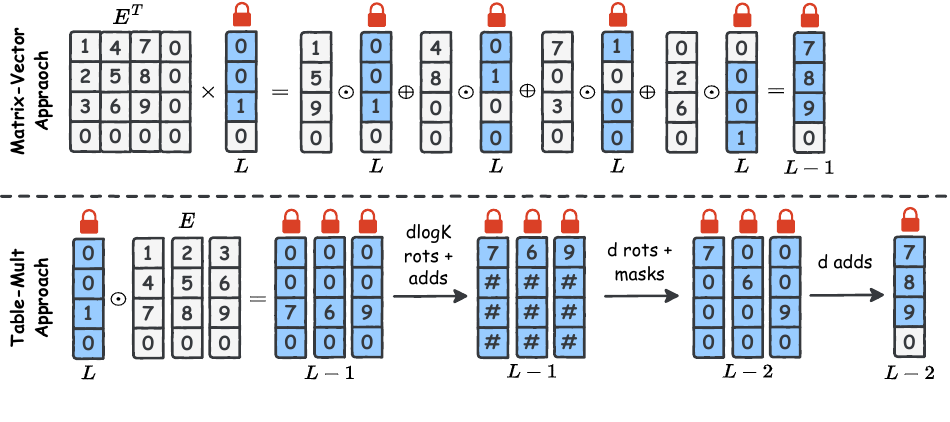}
  \caption{Matrix-vector lookup (top) and column-wise table multiplication (bottom), with $k=d=3$, $n=4$, and $i=2$.
  Both return $(7,8,9)^\top$.
  The first layout consumes one level; the second uses another level to consolidate its $d$ intermediate ciphertexts.
  Here $L$ is the input level, $K=2^{\lceil\log_2 k\rceil}$ is the padded row count, and \# denotes unused slots.}
  \Description{Two circuits select the third row of a three-by-three embedding table. The top circuit multiplies rotated selectors by plaintext diagonals and adds them to obtain a packed vector with entries seven, eight, and nine. The bottom circuit multiplies the selector by each column, reduces each column to a scalar, then masks, rotates, and adds the three ciphertexts into the same packed vector.}
  \label{fig:lookup-formulations}
\end{figure}

Since $s_i$ is encrypted and $E$ is plaintext, the server can evaluate this product using 
CKKS matrix-vector subroutines such as the diagonal Halevi-Shoup method or baby-step giant-step (BSGS) variants~\cite{10.1007/978-3-662-44371-2_31}, with double-hoisted BSGS further reducing rotation cost~\cite{dhbsgs}.
However, this exposes the main inefficiency of representing lookup as dense linear algebra: although the cleartext embedding reads one row of $E$, the homomorphic circuit 
evaluates a full matrix-vector product whose cost scales with the table size.
The plaintext multiplications consume one level.
Figure~\ref{fig:lookup-formulations} (top) shows the diagonal-based approach.

\noindent\textbf{Column-Wise Table Multiplication.}
An alternative applies the encrypted selector vector directly to each column of the embedding table.
Let $E_{:,j}\in\mathbb{R}^k$ denote column $j$ of $E$, for $0\leq j<d$.
The server packs this plaintext column into CKKS slots and multiplies it slot-wise with the selector to obtain $s_i\odot E_{:,j}$.
Since $s_i$ is one-hot, this product is zero in every slot except slot $i$, which contains $E_{i,j}$.
A series of rotations and additions then place the slot sum in the first slot, repeated for each of the $d$ columns.
As shown in Figure~\ref{fig:lookup-formulations} (bottom), masks and rotations place one value from each intermediate ciphertext into its output slot, and adding these ciphertexts forms the packed embedding vector.

The drawback is packing overhead: this layout produces $d$ intermediate ciphertexts that must be consolidated into the packed layout expected downstream, and this masked consolidation consumes an extra multiplicative level relative to the matrix-vector layout.

\subsection{FHE-Efficient Selectors}
\label{sec:ive}

One-hot vectors are just one orthogonal basis for selecting unique embeddings.
Other bases can reduce the cost of constructing the encrypted index vector.
For scalar embeddings $a$ and $b$, the one-hot and dense orthonormal bases retrieve the same entry with different table weights:
\[
\begin{aligned}
  (a,b)(0,1)^\top &= b,\\
  \tfrac12(a+b,a-b)(1,-1)^\top &= b
\end{aligned}
\]
\citet{cheon2026private} use this freedom to construct queries in their Independent Vector Evaluation (IVE) method.

More generally, choose $k$ linearly independent vectors as the columns of $V\in\mathbb{R}^{k\times k}$, with $i$-th column $v_i=Vs_i$ representing index $i$.
Since $V$ is invertible, the server can precompute the plaintext matrix $E^\top V^{-1}$, so that
\[
  (E^\top V^{-1})v_i
  = E^\top V^{-1}Vs_i
  = E^\top s_i.
\]
Under CKKS, the server generates encrypted coordinates of $v_i$ and evaluates this product approximately.

The choice of basis determines the cost of generating $v_i$.
IVE uses an orthogonal discrete cosine transform (DCT) basis.
The client encodes the index as a complex number on the unit circle and encrypts that value.
The server generates its powers through a multiplication tree and uses conjugation to extract their real and imaginary parts.
Together with a constant coordinate, these values form the required DCT column, up to a known scale absorbed into the precomputed matrix.

For power-of-two $k$, vector generation requires $O(k)$ homomorphic operations and $O(\log k)$ multiplicative depth.
Orthogonality gives $V^{-1}=V^\top$, so the basis change preserves the Euclidean norm of errors from vector generation.

IVE therefore accepts one encrypted scalar per lookup and expands it into the coordinates needed to apply the table.
In its batched layout, each slot represents a different query, and each generated ciphertext holds one basis coordinate across those queries.

\subsection{Implications for Industry-Scale DLRMs}
\label{sec:lookup-systems-cost}

Large-scale DLRMs expose the systems cost of private embedding lookup.
Criteo-style recommendation models contain many embedding tables, several of which are much larger than the number of slots ($n$) in a single CKKS ciphertext.
Therefore, the encrypted one-hot selectors 
require many ciphertexts per inference.
For a total of $k_{\mathrm{tot}}$ rows across all tables, concatenating the selectors before encryption requires $\lceil k_{\mathrm{tot}}/n\rceil$ ciphertexts.
For the industry-scale Criteo DLRM with $k_{\mathrm{tot}}\approx33.8\times10^6$ and $n=2^{15}$, this amounts to more than a thousand ciphertexts.

The client-side method must transmit all of these ciphertexts, since sending only the non-zero ones would reveal which table region contains the index.
Thus, even though client-side construction avoids homomorphic work, it can incur substantial communication overhead.
At fixed FHE parameters, reducing the number of used slots reduces the upload only when it changes the number or parameters of the transmitted ciphertexts.

Server-side construction, including IVE, allows the client to send encrypted scalars, but still incurs the cost of vector generation and table multiplication.
These costs must be considered together with the packing needed to assemble the embeddings for the rest of the model.
We address these costs in Section~4 through deterministic compositional embedding compression and a multi-table packing strategy.
Compression reduces the size of the tables; packing combines their lookups into a matrix-vector product whose output is ready for downstream DLRM computation.

\input{04_helrm_design}

\input{05_evaluation}

\input{06_related_work}

\input{07_conclusion}

\bibliography{references}
\bibliographystyle{mlsys2025}

\section*{AI Usage Disclosure}

The authors used Claude Code and Codex during the preparation of this
work to assist with rephrasing and editing text, generating and
refining plots, and developing portions of the code and tests. All AI-assisted outputs,
code, plots, and experimental results were reviewed and validated by
the authors, who take full responsibility for the contents of this work.

\clearpage
\input{10_appendix}
\end{document}

%% file: 00_abstract.tex
\begin{abstract}

FHE inference systems primarily target models with dense inputs, while Deep Learning Recommendation Models (DLRMs) contain sparse categorical features that require lookups into embedding tables.
Encrypted lookups are challenging because indexing is not natively supported and must be expressed using SIMD arithmetic.
A direct one-hot formulation does not scale.
Prior work addresses parts of this problem through embedding compression or compact encrypted queries. However, it does not address complete DLRM inference, leaving open how private lookup interacts with downstream computation and memory constraints. We present HE-LRM, the first end-to-end FHE system for private DLRM inference. We use deterministic compositional embedding compression with client-side digit decomposition, avoiding expensive homomorphic selector construction while consuming only one multiplicative level. We further develop a block-diagonal packing strategy that performs lookups across multiple embedding tables using optimized BSGS linear transformations and produces encrypted embeddings directly compatible with downstream inference. 
We extend Orion with a high-performance GPU backend and achieve 9 ms for single-inference embedding lookup. We also show that GPU memory capacity limits batching and therefore the latency-bound throughput in such systems. On Criteo, HE-LRM achieves complete encrypted inference with server-side latencies of 154–247 ms on a single GPU. We open-source HE-LRM at 
\textcolor{blue}{\href{https://github.com/baahl-nyu/orion}{\fontfamily{lmtt}\selectfont https://github.com/baahl-nyu/orion}}
\end{abstract}

%% file: 01_introduction.tex
\section{Introduction}

Fully Homomorphic Encryption (FHE) enables computation directly on encrypted data~\cite{10.1145/1536414.1536440}. This makes FHE well suited for inference in the cloud, where models regularly process sensitive user data. Recent systems have demonstrated private inference for increasingly complex CNNs and MLPs \cite{orion,gilad2016cryptonets,gazelle,delphi}, but these systems primarily assume dense inputs. Deep Learning Recommendation Models (DLRMs) additionally contain sparse categorical features that are transformed through large learned embedding tables \cite{naumov2019deep}. 
These features encode user-specific information used to personalize recommendations, making DLRMs a compelling target for private inference. Yet end-to-end FHE inference for DLRMs remains largely unexplored.

The challenge is private embedding lookup. In cleartext, a categorical feature simply indexes a row of an embedding table. Under FHE, the index itself is sensitive and must remain encrypted. The CKKS FHE scheme does not provide encrypted indexing; computation is instead expressed using SIMD addition, multiplication, and rotation. A naive solution is to encode each index as an encrypted one-hot vector and evaluate the lookup as a matrix-vector product. Unfortunately, this scales with the number of embedding rows. 
A Criteo DLRM contains 26 categorical embedding tables with a combined 33.8 million rows \cite{criteo}; a direct one-hot formulation requires 1,031 ciphertexts, more than 1 GiB of encrypted input, and more than 1 TiB of server-side FHE plaintext storage. Private embedding lookup therefore introduces substantial communication, computation, and memory costs before the remainder of the network is evaluated.

\begin{figure*}[t]
 \centering
 \includegraphics[width=0.85\textwidth]{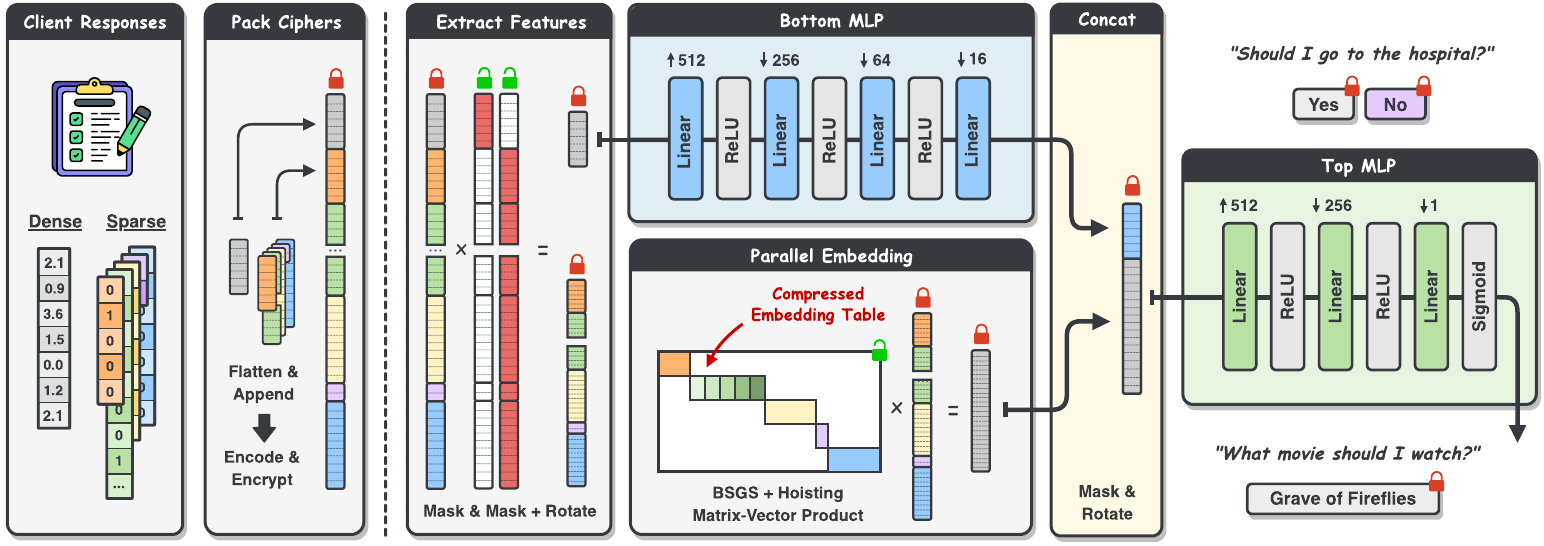}
 \caption[DLRM architecture]{Architecture of a Deep Learning Recommendation Model (DLRM). HE-LRM utilizes Fully Homomorphic Encryption (FHE) to perform end-to-end encrypted inference of this model, maintaining data privacy.}
 \label{fig:dlrm_architecture}
\end{figure*}


Prior work reduces parts of this cost, but exposes different tradeoffs. \citet{kim2024privacypreserving} combine coded embedding compression with server-side encrypted indicator functions, consuming substantial multiplicative depth and server compute. IVE \cite{cheon2026private} instead reduces communication by sending compact encrypted indices and generating an alternative lookup vector homomorphically on the server. This avoids large one-hot uploads, but shifts work to server-side vector generation and relies on batching to amortize that cost. IVE additionally focuses primarily on embedding lookup rather than complete encrypted inference.
For recommendation systems operating under strict latency SLAs \cite{gupta2020deeprecsys,gupta2020architecturalimplicationsfacebooksdnnbased}, the relevant question is therefore not lookup latency or communication in isolation, but latency-bound throughput once downstream computation, batching, and accelerator memory are considered together.

In this work, we introduce HE-LRM, the first end-to-end FHE system for private DLRM inference. First, we develop an FHE-friendly embedding compression technique that maps categorical indices to deterministic coded tokens through client-side digit decomposition. Unlike prior approaches that construct encrypted selectors homomorphically, HE-LRM performs this inexpensive transformation before encryption without requiring any embedding-derived information from the server. The resulting lookup consumes only one multiplicative level and can be evaluated using optimized homomorphic 
transformations.
Second, we address the many heterogeneous embedding tables present in industry-scale DLRMs with a block-diagonal packing strategy that performs multiple lookups in parallel while improving ciphertext slot utilization and remaining compatible with double-hoisted baby-step giant-step (BSGS) \cite{dhbsgs}.
The resulting embeddings are already packed in the layout required by downstream FHE computation.
Together, these techniques provide a CKKS-compatible private lookup primitive that avoids expensive server-side selector construction, does not require exposing embedding-derived mappings or correlations to the client, and composes directly with the rest of the encrypted network.

Beyond the embedding lookup itself, HE-LRM addresses the systems challenges of executing encrypted recommendation inference on GPUs.
We integrate the complete DLRM pipeline into Orion \cite{orion} and extend Orion with a GPU backend built on Cheddar \cite{arxiv-2024-cheddar}, so the Bottom MLP, parallel embedding lookups, concatenation, and Top MLP all execute on one FHE stack on the GPU.
We further exploit unused CKKS slots to batch multiple recommendation queries and amortize FHE computation.
This optimization is particularly important for compact-query approaches such as IVE \cite{cheon2026private}, which trade reduced communication for additional server-side vector generation that can be amortized across queries.
However, larger batches also increase the FHE working set, and GPU memory ultimately bounds the amount of useful amortization.
Because production recommendation systems operate under strict latency constraints \cite{gupta2020deeprecsys,gupta2020architecturalimplicationsfacebooksdnnbased}, we evaluate both latency and latency-bound throughput subject to accelerator memory capacity rather than assuming unconstrained batching.
Finally, we implement IVE \cite{cheon2026private} on the same Orion/Cheddar backend, enabling a direct comparison on common FHE software and hardware.
We open-source HE-LRM together with the Cheddar backend for Orion and our IVE implementation.

Our evaluation shows that these system-level effects materially change the private-lookup tradeoff. At comparable embedding compression of approximately 700$\times$ and batch size one, HE-LRM requires 9 ms for the embedding lookup compared with 165.3 ms for IVE~\cite{cheon2026private}, \textbf{an 18.4$\times$ speedup}. IVE can amortize its additional server computation by batching, but is bounded by accelerator capacity: once the working set exceeds GPU memory, paging sharply increases latency and limits attainable latency-bound throughput. Across Criteo compression configurations, HE-LRM executes the complete encrypted DLRM server-side inference in 88-103 seconds on a single-threaded CPU and 154-247 ms on a single RTX PRO 6000 GPU. These results demonstrate sub-second encrypted recommendation inference and show that private embedding lookup must be designed and evaluated as part of the complete system.

We make the following contributions:

\begin{itemize}
    \item \textbf{End-to-end encrypted DLRM inference.}
    HE-LRM is the first FHE system to run a complete DLRM with dense and sparse inputs, with sub-second inference on Criteo on one GPU.

\item \textbf{A one-level private embedding lookup.}
Client-side digit decomposition compresses embedding tables without homomorphic selector construction and uses a single multiplicative level.


    \item \textbf{Block-diagonal multi-table packing.}
    All embedding lookups run as one BSGS matrix-vector product whose output is in the layout the rest of the DLRM needs.

    \item \textbf{An open-source GPU FHE stack.}
    We add a Cheddar GPU backend to Orion, implement IVE on it, and open-source both with HE-LRM.

    \item \textbf{Memory-bound throughput analysis.}
    Across four GPUs, we show how batching, compression, and GPU memory jointly determine latency-bound throughput.
\end{itemize}

%% file: 04_helrm_design.tex
\section{HE-LRM Design}
\label{sec:helrm_design}
As shown in the previous section, conventional approaches to private embedding lookup under FHE scale poorly, especially for Criteo-sized DLRMs with many large embedding tables.
In this section, we first introduce a deterministic compositional embedding compression technique that extends well-known embedding compression methods that do not affect our threat model: the client need only know the size of these compressed tables to perform private two-party inference.
Second, we formulate compressed private embedding lookup as a plaintext-matrix / encrypted-vector multiplication, enabling efficient evaluation with baby-step giant-step (BSGS) algorithm.
We further extend this formulation with a multi-table packing strategy that evaluates several DLRM embedding lookups in a single packed matrix-vector product, reducing both computational and memory pressure.
Third, we apply standard CKKS optimizations and do batched inference to measure the forward pass latency and meet the latency-bound throughput SLAs.
We also propose a fast, open source implementation of the IVE selector that can run on the Orion's \cite{orion} GPU backend using Cheddar \cite{arxiv-2024-cheddar} and compare against it.


\begin{figure}[!t]
\centering
\includegraphics[width=0.872\columnwidth,trim=4.56 3.96 3.48 1.68,clip]{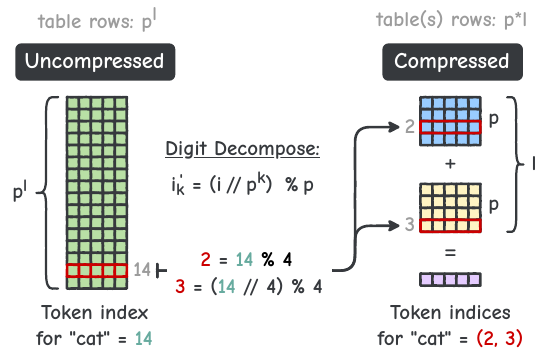}
\caption{Our compressed embedding lookup technique. Client-side base-$p$ digit decomposition maps token ``cat'' (14) to coded tokens $(x_0=2, x_1=3)$. Each coded token then retrieves a row from the corresponding compressed tables which are then added together.}
\label{fig:our_optimizations}
\end{figure}

\subsection{Deterministic Base-$p$ Digit Decomposition}
\label{subsec:digit_decomposition}

We build on the Quotient--Remainder (QR) method of \citet{shi2020compositional} to efficiently perform private embedding lookups.
For an embedding table with $k$ rows, QR maps an input index $i$ to the pair
$(\lfloor i/q \rfloor,\, i \bmod q)$ for some $q < k$.
The two resulting indices address separate embedding tables whose outputs are combined, and the tables are trained end-to-end.

We generalize this idea from two components to $\ell$ base-$p$ digits.
For an input index $i \in \{0,\ldots,k-1\}$, we choose
$\ell = \lceil \log_p k \rceil$, so that $p^\ell \geq k$, and decompose $i$ into
$(i_0,\ldots,i_{\ell-1})$, where each $i_j \in \{0,\ldots,p-1\}$.
The compressed embedding tables are learned jointly with the rest of the DLRM, reducing the embedding parameter count from $d p^{\ell}$ to $d p \ell$.
Because the digit decomposition is deterministic, it requires no mapping table for the client to obtain or the server to store.

Figure~\ref{fig:our_optimizations} illustrates the decomposition for an embedding table with $k=16$ rows and embedding dimension $d=5$.
Choosing $p=4$ gives $\ell=\lceil \log_4 16\rceil=2$ digits.
For example, the index $i=14$ decomposes as $(i_0,i_1)=(2,3)$ because
$14 = 2\cdot 4^0 + 3\cdot 4^1$.
Each digit selects one row from its corresponding compressed embedding table, and the selected vectors are summed to produce the final $d$-dimensional embedding.

\subsection{Packing Multiple DLRM Lookups}
\label{subsec:multiple_tables}

\noindent\textbf{DLRMs require many embedding tables.}
Our discussion in the previous sections was limited to performing an encrypted lookup into a single (compressed or uncompressed) embedding table, but as Section~\ref{sec:lookup-systems-cost} shows, industry-scale DLRMs have many embedding tables of different sizes.
Under the constraints of FHE, we are posed with the following question: how can we efficiently perform encrypted lookups when we have multiple embedding tables?

A straightforward solution to this multi-embedding table case is to simply encrypt each compressed index \textit{separately} and perform the methods discussed before. However, this method would require one CKKS ciphertext per embedding table and may be wasteful when compressed tables have a total size less than the number of available CKKS slots.

\begin{figure}[!t]
\centering
\includegraphics[width=0.9\columnwidth]{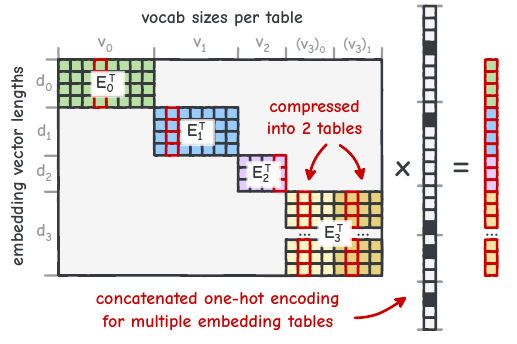}
\caption{Block-diagonal packing of multiple embedding tables into a single matrix-vector product, supporting both compressed and uncompressed tables while producing contiguous output embeddings.}
\label{fig:block_diagonal}
\end{figure}

\noindent\textbf{Block-diagonal embedding matrix.}
To this end, we propose a multi-embedding packing strategy which places unique embedding tables \textit{diagonally} across a standard weight matrix by starting each subsequent embedding table at the bottom right corner of the previous embedding table. This packing strategy is akin to calling: \texttt{torch.block\_diag} on the list of embeddings. Importantly, the block-diagonal packing strategy is interoperable with both compressed and uncompressed embedding tables. An example of our packing strategy can be seen in Figure~\ref{fig:block_diagonal} where four feature embedding tables are stored in a single weight matrix. The client simply one-hot encodes each index for each table (again, either compressed or uncompressed) and concatenates these one-hot vectors into a single vector, which can then be encrypted.

The resulting matrix-vector product exactly extracts and combines the correct columns of the packed embedding matrix and places the encrypted embedding vectors into contiguous slots within an output ciphertext. Importantly, this diagonal multi-embedding packing strategy aligns well with diagonal-based BSGS matrix-vector product strategies used in prior work \cite{orion, garimella2025einhops}. Moreover, diagonally aligning embedding tables incurs little FHE overhead as long as the sum of the embedding dimensions does not exceed the slot count.


\subsection{HE-LRM Protocol}
\label{subsec:protocol}

\begin{figure}[!t]
\centering
\includegraphics[width=0.9\columnwidth]{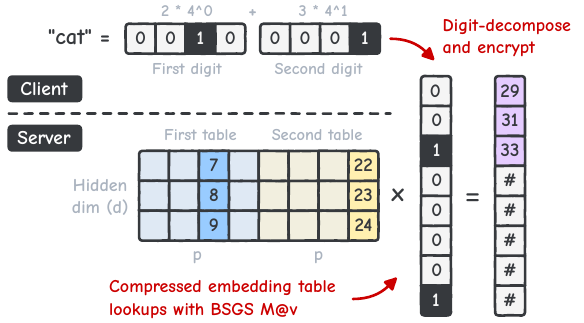}
\caption{Our proposed solution for compressed embedding lookups. The client deterministically decomposes each token index into base-$p$ digits, constructs the corresponding one-hot selectors locally, and encrypts them. The server
packs compressed embedding tables so all lookups reduce to a BSGS matrix-vector product whose output is already arranged for the following inference.}
\label{fig:our_pipeline}
\end{figure}

Figure~\ref{fig:dlrm_architecture} shows the end-to-end HE-LRM inference pipeline, and Figure~\ref{fig:our_pipeline} details its embedding lookup.
The client deterministically decomposes each categorical index into base-$p$ digits and locally constructs the corresponding one-hot encodings.
It concatenates these with its dense features, encrypts the result under CKKS, and uploads it during offline ingestion, making the encrypted query available on the server before inference.
Because the client constructs and encrypts the selectors, the server never learns the index or evaluates an expensive encrypted indicator function.
Based on the DLRM configuration, the server homomorphically extracts the packed dense and sparse components, sending the dense features through the Bottom MLP and using the sparse one-hot features for embedding lookups.
For the lookups, we concatenate and transpose the compressed embedding tables into the block-diagonal matrix of Section~\ref{subsec:multiple_tables}, enabling a single plaintext-matrix/encrypted-vector multiplication.
The existing BSGS matrix-vector routine applies directly, producing encrypted embeddings already packed for the rest of the model.
The server concatenates the Bottom MLP and embedding outputs, passes the resulting ciphertext through the Top MLP, and returns an encrypted logit to the client service.


\subsection{Batched Inference with Memory Ceiling}
\label{subsec:batching}
CKKS cost trends with the amount of ciphertexts, regardless of the slot-occupancy of these ciphertexts. A ciphertext having $N/2$ slots needs to be processed the same way regardless of how many slots have meaningful data. Therefore, in cases where the query occupies less than $N/4$ slots, multiple queries can be batched together in the same ciphertext. 

In DLRMs, user data along with impressions are already present on the server and need to be batch processed.
Since HE-LRM's encrypted queries are likewise resident on the server (Section~\ref{subsec:protocol}), multiple queries can be packed in the same ciphertext given enough slots.
This amortization benefits both IVE as well as our approach by different margins.

However, in a heterogeneous system, we must consider that batching multiple queries has an impact on peak memory usage. GPUs have a fixed memory ceiling, and going off chip severely downgrades the performance of FHE kernels. Therefore, we encounter a situation where the batch size not only determines the throughput, but also depends on the GPU memory capacity. Figure~\ref{fig:memory-vs-batch} shows how memory climbs with various batch sizes and compressions for our and IVE’s approach. Even at the highest compression in both approaches, batches beyond size 16 spill over our GPU memory ceiling of 96 GiB and become infeasible for latency-bound throughput due to memory paging costs.
Going beyond the memory ceiling is detrimental to the latency-bound throughput regardless of what method is used to perform encrypted DLRM inference as shown in Section~\ref{sec:results}.

%% file: 05_evaluation.tex
\begin{figure}[t]
\centering
\includegraphics[width=0.94\columnwidth]{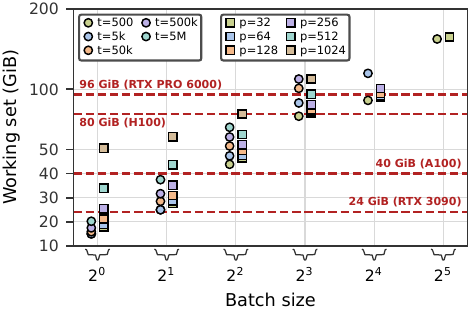}
\caption{Peak GPU working-set memory versus batch size for HE-LRM and IVE across compression settings. Dashed lines mark the memory capacities of the evaluated GPUs; configurations above these limits no longer fit entirely in
device memory.}
\label{fig:memory-vs-batch}
\end{figure}

\section{End-to-End Results for DLRM inference}
\label{sec:results}

\subsection{Setup}
\label{subsec:setup}

We conduct our CPU experiments on an AMD Ryzen Threadripper 9960X, 24 cores processor and 256 GiB of RAM. Our GPU experiments run on a single RTX PRO 6000 Blackwell Workstation Edition with 96 GiB of GPU RAM; unless stated otherwise, all GPU results are on this card. We also evaluate our throughputs (Section~\ref{subsec:ive}) on an NVIDIA A100 (40 GiB), an NVIDIA H100 (80 GiB) and an RTX 3090 (24 GiB). We train all DLRMs on the RTX PRO 6000, and experimental results (training losses, AUCs, and FHE latencies) are all averaged over 3 runs.

Our implementation builds upon the CAFE framework~\cite{liu2025cafe+}, which provides an optimized DLRM implementation.
We implement the generalized QR decomposition technique within CAFE.
We use the hyperparameters set by their repository.
We port all DLRM models directly into the Orion framework and implement all necessary components (extraction, concatenation, embeddings).
Unless stated otherwise, we choose an FHE parameter set that enables bootstrapping while maintaining 128-bit security.
We collect GPU results with our Cheddar backend for Orion, which uses 32-bit words with rational rescaling \cite{arxiv-2024-cheddar}.
We also implement the IVE selector in Orion’s Cheddar backend and use it in the HE-LRM pipeline to assert how GPU memory limits the amortization benefits of this approach.
As described in Section~\ref{subsec:protocol}, the encrypted queries are resident on the server, so our system model does not include network cost and we measure the latency of the server forward pass.
Appendix~\ref{subsec:network} analyzes the upload cost of a live query separately.


We evaluate HE-LRM on the Criteo 7-day click-through dataset. It includes 13 continuous features (normalized integers) and 26 categorical features where each feature has been anonymized. There are approximately 45 million training samples and this dataset is used to predict the click-through-rate for a provided advertisement. The first 6 days of data are used for training while the seventh day serves as the validation set.

\subsection{Training DLRMs}
\label{subsec:training}

We train DLRMs on Criteo using CAFE.
Every embedding table with more than $t$ rows is compressed with the base-4 digit decomposition of Section~\ref{subsec:digit_decomposition}, and smaller tables are left uncompressed.
Similar to prior work~\cite{liu2025cafe+}, we define the Compression Ratio as the ratio of the total number of embedding rows in the full DLRM to the total number embedding rows in a compressed model. We vary the threshold to roughly track an order of magnitude of increases in the compression ratio. Table~\ref{tab:cpu_vs_gpu} shows our chosen thresholds and their exact CR. A low threshold (500 rows) compresses a majority of the 26 sparse features for Criteo and reduces the entire embedding size by $31180\times$. 
Appendix~\ref{subsec:appendix_training} provides more details about the training runs.


\subsection{HE-LRM on CPU against GPU}
\label{subsec:results}


Table~\ref{tab:cpu_vs_gpu} compares our CPU and GPU performance at various thresholds for batch size 1.
Latency grows with the number of ciphertexts in the embedding lookup, and the three most compressed models, which use 1--2 ciphertexts, perform similarly.
This is because apart from the embedding table lookups, the DLRM architecture is the same across thresholds, meaning that all FHE operations are identical outside of the embedding lookup.
At highly compressed models, the entire query fits in a single ciphertext and the lookup is processed within 1-2s on the CPU and 4-9ms on the GPU.
On the other hand, less compressed models require more ciphertexts (18 for threshold 500k and 85 for threshold 5m), inflating the matrix-vector product latency; for example the largest model takes 15.2 seconds to perform the embedding lookup on the CPU and 53ms on the GPU.
Since GPU inference is 417--573$\times$ faster than CPU inference across the evaluated compression thresholds, we focus the remaining evaluation on GPU throughput under latency constraints.

\begin{table}[t]
\centering
\small
\caption{HE-LRM inference latency at batch size 1 across compression
thresholds. CTs denotes the number of ciphertexts used by the compressed
embedding lookup}
\label{tab:cpu_vs_gpu}
\vskip 0.15in
\setlength{\tabcolsep}{3.5pt}
\begin{tabular}{r r c c c c}
\toprule
Threshold & Compression & CTs & CPU (s) & GPU (s) & Speedup \\
\midrule
500       & 31,180$\times$ & 1  & 88.2  & 0.154 & 573$\times$ \\
5,000     & 3,745$\times$  & 1  & 88.2  & 0.154 & 572$\times$ \\
50,000    & 707$\times$    & 2  & 89.4  & 0.159 & 561$\times$ \\
500,000   & 59.3$\times$   & 18 & 94.7  & 0.183 & 519$\times$ \\
5,000,000 & 12.2$\times$   & 85 & 103.2 & 0.247 & 417$\times$ \\
\bottomrule
\end{tabular}
\vskip -0.1in 
\end{table}

\subsection{Comparison with IVE}
\label{subsec:ive}

We implement IVE in Orion using the same Cheddar backend as HE-LRM, so both selectors run on the same FHE stack. IVE natively supports lookups from a single embedding table. To support lookups across tables, we apply our block-diagonalization method from Section~\ref{subsec:multiple_tables} to IVE.

\begin{figure}[t]
\centering
\includegraphics[width=0.94\columnwidth]{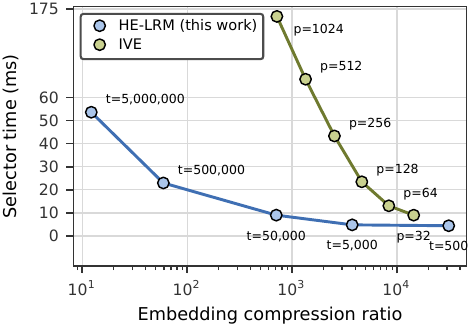}
\caption{GPU selector latency at batch size 1 versus embedding compression
ratio ($d=16$). At the closest matched compression, HE-LRM requires
9.0 ms at $707\times$ compression, compared with 165.3 ms for IVE at
$717\times$, an $18.4\times$ speedup.}
\label{fig:selector-cost}
\end{figure}

\noindent\textbf{Selector cost.}
We compare both selectors at batch size~1. Figure~\ref{fig:selector-cost} shows the GPU selector time of both designs against compression. The rest of the forward is the same ($\approx$150 ms) for both designs, so the selector gap becomes a 2.0$\times$ end-to-end advantage at 700$\times$ compression (159 against 314 ms) and a 3--10\% one above 4{,}000$\times$. Additionally, we provide a detailed stage-wise breakdown of the selector costs in Appendix~\ref{subsec:appendix_selector_cost}.

IVE recovers this gap by amortization. Its 432-slot query packs up to 75 queries into one ciphertext, so the selector cost is flat in the batch and the per-query cost falls as $1/B$. HE-LRM batches differently: it concatenates the one-hot digits of consecutive queries along the slot axis and packs them densely across ciphertexts. At $t{=}50{,}000$ a query spans 47{,}747 slots, so batches of 1, 2 and 4 use 2, 3 and 6 ciphertexts and the BSGS lookup grows with that count; at $t{=}5{,}000$ a query spans 9{,}015 slots and three share a ciphertext, and at $t{=}500$ it spans 1{,}083, so a batch of 16 is still a single ciphertext. Either way, the MLPs run once on the packed ciphertext, so for both designs the forward at batch $B$ costs little more than at batch 1 and throughput grows almost linearly with $B$. The question is therefore how large $B$ can be, and that is set by memory, not by the selector.

\begin{figure}[t]
\centering
\includegraphics[width=0.9\columnwidth]{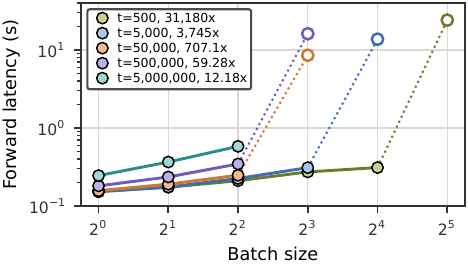}
\caption{HE-LRM inference latency versus batch size on the RTX PRO 6000 across
compression thresholds. Filled markers denote batches that fit in GPU memory. The hollow marker for each configuration is the first batch whose working set exceeds the 96 GiB device capacity.}
\label{fig:latency-vs-batch}
\end{figure}

\noindent\textbf{Off-chip penalty.}
Figure~\ref{fig:latency-vs-batch} shows HE-LRM inference latency versus batch size across compression thresholds on the RTX PRO 6000. For each configuration, we include the first tested batch size whose working set exceeds GPU memory. We run these batches with CUDA managed memory, which pages data between host and device memory.
At the largest batch sizes that fit in GPU memory, HE-LRM inference takes 0.25--0.58~s. At the next tested batch sizes, latency increases to 8.6--24~s. IVE exhibits the same pattern, with a 27--35$\times$ latency increase when its working set exceeds GPU memory. These results show that paging sharply increases inference latency for both methods. Even the largest batches that fit in memory exceed the 66--150~ms plaintext production targets reported by DeepRecSys~\cite{gupta2020deeprecsys}. This sharp increase in latency suggests that encrypted recommender systems should keep batches small enough to fit entirely on the GPU, rather than optimizing for throughput.


\noindent\textbf{Memory capacity vs. throughput.}
Figure~\ref{fig:memory-vs-batch} shows peak GPU memory usage across batch sizes and compression ratios.
These measurements identify the largest batch that fits entirely in memory on each GPU.
We evaluate throughput at this batch size, dividing the number of queries by the batch's inference latency.
GPU memory capacity therefore places a practical limit on throughput gains from batching.

\begin{figure}[t]
\centering
\includegraphics[width=0.9\columnwidth]{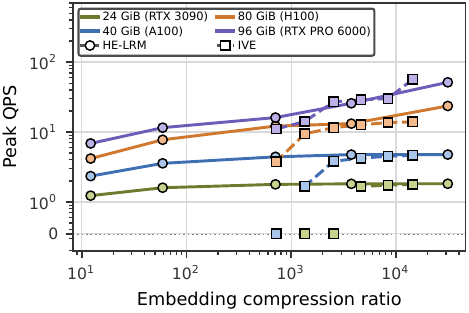}
\caption{Peak throughput versus embedding compression ratio for HE-LRM and IVE across four GPUs. Throughput is measured at the largest batch size that fits entirely on device.}
\label{fig:peak-qps}
\end{figure}

\noindent\textbf{Peak throughput across GPUs.}
Figure~\ref{fig:peak-qps} compares throughput across four GPUs using the largest batch sizes that fit in GPU memory. HE-LRM fits on all four GPUs at every evaluated compression threshold. IVE exceeds memory capacity for $p>256$ on the RTX 3090 and for $p=1024$ on the A100. For the remaining settings on these two GPUs, IVE fits but supports smaller batches than HE-LRM.

As GPU memory capacity increases, IVE can process larger batches and reduce its selector cost per query. The throughput gap therefore narrows on the RTX PRO 6000, where the two methods achieve similar throughput at the highest compression settings. IVE reaches 56.8 queries per second at $p=32$, compared with 51.4 for HE-LRM. On the H100, HE-LRM retains an advantage because it supports a larger batch, achieving 23.6 queries per second compared with 14.0 for IVE. At lower compression ratios, below approximately $1{,}500\times$, HE-LRM achieves higher throughput on all four GPUs. Differences in inference latency also affect throughput. Using the RTX PRO 6000 as the reference, inference latency is $1.5\times$ on the H100, $2.5\times$ on the A100, and $3.5\times$ on the RTX 3090.

%% file: 06_related_work.tex
\section{Related Work}

In a client–server model, the model weights remain hidden from clients and inputs from the server. Private inference protocols commonly use FHE and evaluate the whole network server-side, approximating nonlinear layers with low-degree polynomials, trading some accuracy for performance. Early systems (CryptoNets, MiniONN, GAZELLE, DiNN) demonstrated feasibility~\cite{gilad2016cryptonets,liu2017oblivious,juvekar2018gazelle,bourse2018fast}. For more complex tasks such as natural language processing or recommendation systems, embedding layers with large LUTs are a central challenge. In CKKS, prior work avoids large encrypted LUTs by (i) sending high-dimensional encrypted one-hots~\cite{al2020privft} or (ii) computing embeddings in plaintext on the client~\cite{lee2022privacy,chen2022x}. Both solutions have limitations with proprietary tables or tight bandwidth. \citet{kim2024privacypreserving} enable efficient encrypted evaluation of large LUTs in CKKS with compression.
IVE~\cite{cheon2026private} reduces communication with compact encrypted indices at the cost of additional server-side computation.

\subsection{Private information retrieval}
\noindent Standard private information retrieval (PIR) allows a client to retrieve a database entry without revealing the accessed index. This approach differs from those in DLRMs, which require further server-side computation on the resulting table entry. We note that unlike CKKS-based private language models, private inference protocols which employ multiple cryptographic primitives (e.g., multi-party computation and BFV fully homomorphic encryption) do perform encrypted embedding lookups \cite{ciphergpt, lin2024}. Recent work \cite{kim2026ive, ji2026gpir} has also accelerated PIR with dedicated ASICs and on GPUs.

\subsection{Embedding compression for DLRMs}

Embedding tables in DLRMs can dominate memory and bandwidth which bottlenecks inference latency. Prior work compresses these tables to cut storage and compute while preserving accuracy. For instance, quantization-based methods use lower-precision data types to reduce memory footprint through methods like uniform or row-wise quantization, applied during training ~\cite{zhou2024dqrm,kang2020learning,ko2021mascot} or post training~\cite{guan2019post,deng2021low}. Clustering-based methods can also represent embeddings with shared centroids for high compression~\cite{shu2018compressing}.
Hashing-based methods (e.g., quotient-remainder decomposition ~\cite{shi2020compositional}) map large input IDs into a smaller embedding space, trading memory for collision risk~\cite{weinberger2009feature}. Hashing is often combined with the aforementioned methods~\cite{tsang2023clustering,ghaemmaghami2022learning}.
Adaptive compression methods dynamically allocate memory based on feature importance. One example is CAFE+~\cite{liu2025cafe+}, which tracks the importance of each feature and combines this dynamic strategy with hashing and quantization to scale to very large models.





\subsection{FHE compilers and accelerators}

\noindent \textbf{Compilers.}
Compiler for FHE can be divided into two categories: circuit-level compilers~\cite{heco, lee2022hecate, cowan2021porcupine,malik2023coyote,lee2023elasm}, focusing on low-level optimization and scheduling for general-purpose programs, and
domain-specific compilers (early works include~\cite{dathathri2019chet, dathathri2020eva}).
More recent compilers Dacapo~\cite{cheon2024dacapo}, HeLayers~\cite{aharoni2020helayers}, Fhelipe~\cite{krastev2024tensor},
 Orion~\cite{orion} improve upon these works by introducing automating bootstrap placement and support for deeper models.


\noindent \textbf{Accelerators}:
cuHE~\cite{cuHE} was among the first works to accelerate FHE on GPUs, paving the way for subsequent GPU implementations such as~\cite{gpuboot} and Cheddar~\cite{arxiv-2024-cheddar}.
FPGAs~\cite{heax, agrawal2022fab, poseidon} and custom ASICs~\cite{feldmann2021f1, reagen2020cheetah, craterlake, bts, ark, sharp, 10992987, ebel2024osiris} have also been used to accelerate FHE, with ASICs improving performance by several orders of magnitude.
However, ASICs have high design and fabrication costs and limited flexibility, which makes them harder to deploy than commodity GPUs.

%% file: 07_conclusion.tex
\section{Conclusion}
We present HE-LRM, the first end-to-end private DLRM under FHE. On Criteo, HE-LRM performs a complete encrypted inference in 88--103 s on one CPU thread and 154--247 ms on one GPU. Our evaluation shows that GPU memory capacity limits batch size for low-latency private recommendation, and HE-LRM's smaller memory footprint per query supports larger batches and enables inference across GPUs.
The embedding lookup problem extends to other applications such as language models, which need batched queries on sparse-embeddings. In an encrypted setting, the performance and feasibility of these models will depend on the batch size, system constraints, and the service requirements. Overall, our results motivate co-designing encrypted data representation, computation, batching, and accelerator memory resources for sparse-embedding workloads under~FHE.

%% file: 10_appendix.tex
\newpage
\appendix
\section{CKKS operations}
\label{subsec:appendix_ckks}

CKKS~\cite{cheon2017homomorphic, ckks2} is a SIMD style homomorphic encryption scheme that encrypts a vector of complex or real values into a ciphertext. 
This scheme relies on RingLWE ciphertexts in ${\mathcal{R}_Q^2}$ for a given ring $\mathcal{R}=\mathbb{Z}[X]/(X^N+1)$, where $N = 2^k$, for $k > 0$, is the ring dimension and $\mathcal{R}_Q=\mathcal{R}/Q\mathcal{R}$ describes the ring $\mathcal{R}$ reduced modulo an integer $Q = \prod_{i=0}^{L}q_{i}$. 
The integer $L$ is known as the multiplicative depth and represents the maximum number of rescaling levels available before decryption fails.
CKKS supports element-wise operations such as addition and multiplication as well as cyclic rotation.
Based on the characteristics of the CKKS scheme, it becomes a natural choice for applications such as deep learning recommendation models where real-valued vectors are considered.
We now describe some of the core operations used with CKKS.

\noindent \textbf{Encoding.} Consider a real (or complex) vector $\textbf{x} \in \mathbb{C}^{N/2}$. This vector can be encoded into elements of $\mathcal{R}_Q$ using an approximate inverse of a scaled complex canonical embedding. 
More precisely, one applies an inverse Fast Fourier Transform on the elements of $\textbf{x}$ and scales each output by a scaling factor $\Delta$. Finally, each element is rounded to the nearest integer as encryption is performed over integers modulo $Q$.
This step outputs a plaintext polynomial $\textbf{m}(X)$ which 
packs $N/2$ complex values. These $N/2$ values are referred to as the available \textit{slots} in a CKKS object.

\noindent\textbf{Encryption} The plaintext polynomial $\textbf{m}(X)$ can be encrypted into a ciphertext $(\textbf{a}, \textbf{b}) \in \mathcal{R}^2_Q$ with a given public key and the addition of some random noise.

\noindent\textbf{Addition.} CKKS supports element-wise plaintext-ciphertext and ciphertext-ciphertext additions. The resulting ciphertext after this operation corresponds to the SIMD addition of the underlying complex vectors.

\noindent\textbf{Multiplication.} Similar to addition, CKKS supports multiplication between either a plaintext and ciphertext or between two ciphertexts. 
Multiplication is followed by a rescaling procedure to avoid an exponential growth in the scaling factor. The resulting polynomial after this operation has coefficients in $\mathbb{Z}_{Q_{\ell -1}}$ instead of $\mathbb{Z}_{Q_{\ell}}$ where $Q_{\ell} = \prod_{i=0}^{\ell}q_i$ for $0 < \ell \leq L$ which corresponds to the consumption of a level in the chain of moduli.

\noindent \textbf{Rotation.} Cyclic rotations shift the elements of the input vector $\textbf{x}$ by $0 < k < N/2$ slots. The resulting ciphertext after this operation corresponds to the same underlying vector with shifted elements by $k$ slots.

\noindent \textbf{Key-switching.} Key-switching is a standard operation in CKKS which converts a ciphertext encrypted under one secret key into a ciphertext that decrypts correctly under another secret key. This procedure is necessary after certain operations such as ciphertext-ciphertext multiplication or ciphertext rotation so that the result remains compatible with the decryption key. The transformation is done using a special switching key, which allows this change without decrypting the ciphertext.

\noindent \textbf{Bootstrapping.} In CKKS, a ciphertext can be \textit{bootstrapped} to  increase its number of available levels. This operation is necessary after the moduli chain has been depleted ($\ell = 0)$ via multiplications and no more levels are available for further computations. Among all CKKS operations, bootstrapping is the most computationally expensive, taking roughly 20 seconds on a single-threaded CPU.

\begin{figure}
  \centering
  \includegraphics[width=\columnwidth]{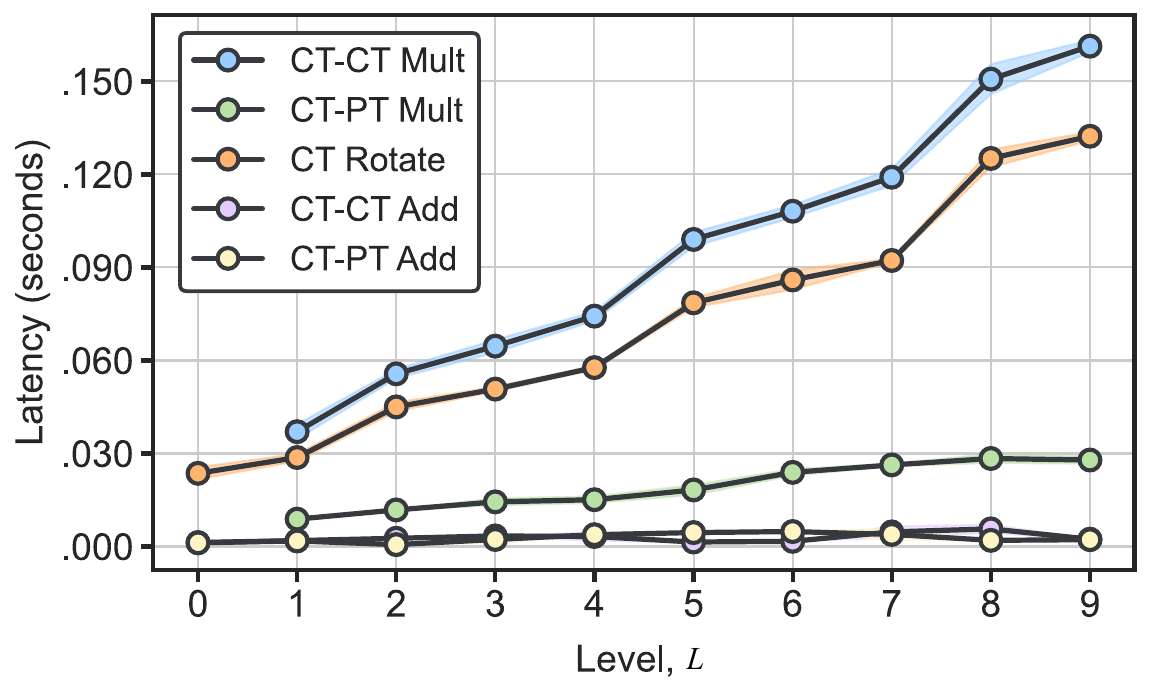}
  \caption[Lattigo Latencies]{Single-threaded CPU latencies of primitive CKKS operations averaged over 30 runs. Both ciphertext-ciphertext multiplication (CT-CT Mult) and ciphertext rotation (CT Rotate) require a compute and memory intensive key-switching operation. CKKS multiplies require ciphertexts to have at least two remaining levels.}
  \label{fig:primitives}
\end{figure}


\begin{figure}
\centering
\includegraphics[width=\columnwidth]{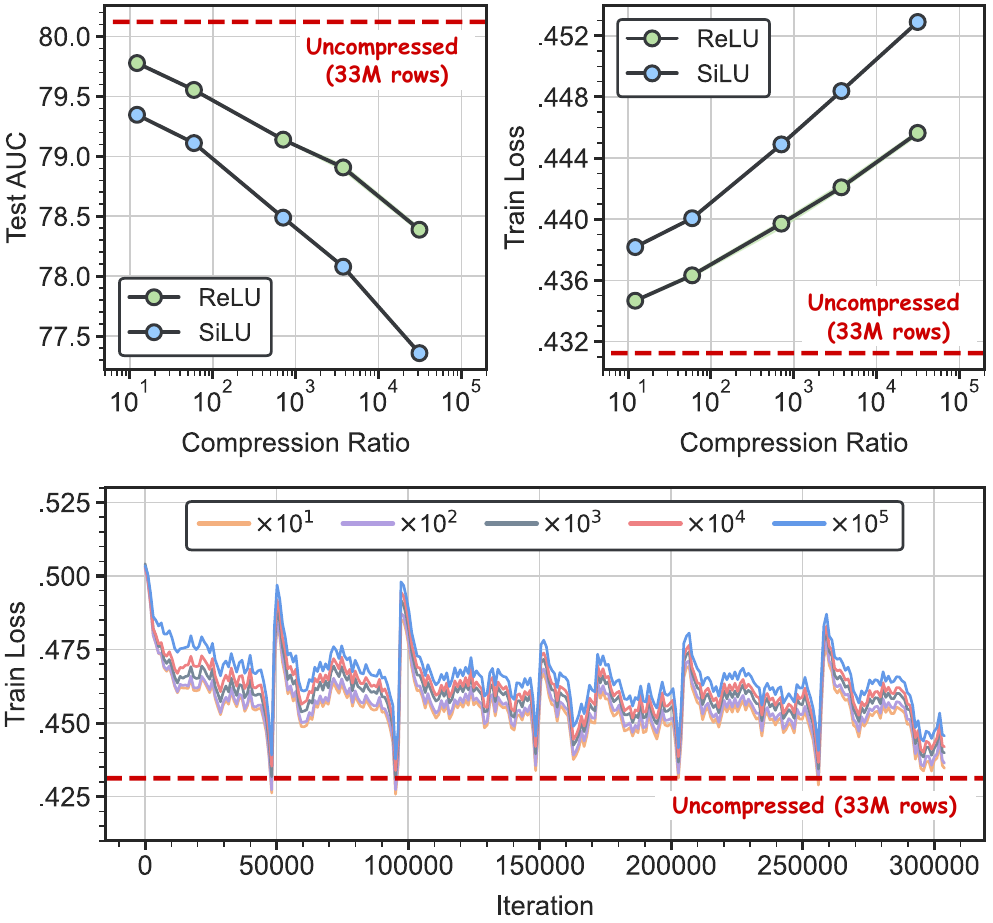}
\caption{Training curves and performance trade-offs for compressed
DLRM models across different compression ratios.}
\label{fig:training_results}
\end{figure}

\section{Training Details}
\label{subsec:appendix_training}

Figure~\ref{fig:training_results} (top) shows test AUC and final training loss against compression for the ReLU and SiLU models; both degrade smoothly and monotonically with compression.

Figure~\ref{fig:training_results} (bottom) displays training runs for each compression threshold configuration, We find that all models follow similar training dynamics and fall naturally on the loss spectrum based on the compression ratio: a less (more) compressed model has a slightly lower (higher) loss over the entire training run. The diurnal trend in the loss is expected: the Criteo dataset is processed in-order over the course of 6 days for training with each new day initially increasing the loss before falling.

We also train each DLRM configuration with the SiLU activation function rather than the standard ReLU activation function. We do this since SiLU is a smoother activation function and therefore is easier to approximate in FHE using a composition of polynomials when compared to ReLU. However, we find that for each configuration, the SiLU networks exhibit both lower test AUC and higher training losses.

\section{Selector Cost Breakdown}
\label{subsec:appendix_selector_cost}
Table~\ref{tab:selector-breakdown} breaks down the selector cost by stage at the closest compression the two parameter sets (HE-LRM and IVE) have. IVE spends 105.7 ms in the power tree, which is used to generate their one-hot-equivalent query, and 56.6 ms in the PC-MM at $p{=}1024$, and takes 165.3 ms in total for the lookup. HE-LRM's BSGS lookup at $t{=}50{,}000$ takes 8.6 ms, 9.0 ms with the repack, an 18.4$\times$ difference. The gap narrows as compression rises in Figure~\ref{fig:selector-cost}, because IVE's power tree shrinks with $p$ while our lookup is already a single ciphertext: at $p{=}32$ against $t{=}500$ the selectors are 9.0 and 4.4 ms. 

\begin{table}[t]
\centering
\small
\setlength{\tabcolsep}{6pt}
\caption{GPU selector cost by stage at batch size 1 for HE-LRM and IVE at closest compression ratio (embedding dimension $d{=}16$).}
\label{tab:selector-breakdown}
\vskip 0.15in
\begin{tabular}{l r r}
\toprule
& \textbf{HE-LRM (ours)} & \textbf{IVE} \\
Operation & $t{=}50{,}000$ & $p{=}1024$ \\
\midrule
Compression ratio          & 707$\times$ & 717$\times$ \\
\midrule
Power tree (VecGen)        & --  & 105.7 \\
PC-MM                      & --  & 56.6 \\
Sub-table sum              & --  & 2.2 \\
BSGS lookup                & 8.6 & -- \\
Repack to top-MLP layout   & 0.4 & 0.9 \\
\midrule
\textbf{Selector total}    & \textbf{9.0} & \textbf{165.3} \\
\bottomrule
\end{tabular}
\end{table}

\section{Network cost of a live query}
\label{subsec:network}
\begin{figure}[t]
\centering
\includegraphics[width=0.94\columnwidth]{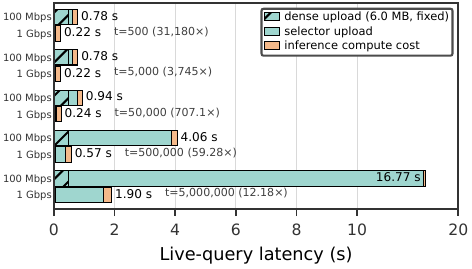}
\caption{Live-query latency for HE-LRM at batch size 1 over 100 Mbps and 1 Gbps network bandwidths. Each bar includes the fixed 6.0 MB dense-feature upload, the compressed selector upload, and the server-side encrypted forward pass.}
\label{fig:network}
\end{figure}
In our system model, users upload encrypted features during an offline ingestion phase, so network transfer is excluded from server-side inference latency. We separately evaluate live queries that require an upload before inference. We process these queries individually, using batch size 1.

Each query upload contains dense features and an embedding selector. The dense features occupy one ciphertext, a fixed upload of 6.0~MB. The selector upload depends on the compression threshold and the ciphertext level required for lookup. HE-LRM uploads vectors at the level required by the embedding lookup. At thresholds of 500 and 5,000 rows, the selector occupies one ciphertext totaling 1.8~MB. At 50,000 rows, it occupies two ciphertexts totaling 3.7~MB. At thresholds of 500,000 and 5,000,000 rows, the selectors require 18 and 85 ciphertexts, totaling 42 and 201~MB.

Figure~\ref{fig:network} shows live-query latency, including upload time and server-side inference, at bandwidths of 100~Mbps and 1~Gbps. For the three most compressed models, uploading the query adds approximately 40--50\% to inference latency at 1~Gbps, with dense features accounting for most of the upload. Less compression increases the selector upload size. For the two least compressed models, selector upload dominates: total latency reaches 0.57 and 1.90~s at 1~Gbps, and 4.1 and 16.8~s at 100~Mbps. If maintaining accuracy requires less compression, uploading the query can therefore take longer than running inference.

IVE also packs its query into one ciphertext, but at comparable compression its upload is larger. It also requires an input ciphertext with $\log_2(p/2)$ levels above what is required by the Top MLP to generate its selector, described in Section \ref{sec:ive}. Together with its higher inference latency, the larger upload gives IVE higher live-query latency at both evaluated bandwidths.